\documentclass[12pt]{article}
\usepackage[english]{babel}
\usepackage{epsfig}
\usepackage{amssymb}
\usepackage{amsmath}%
\usepackage{amsfonts}%
\usepackage{graphicx}
\providecommand{\U}[1]{\protect\rule{.1in}{.1in}}
 0

\begin{document}

\title{Quantum corrections to static solutions of Nahm equation and Sin-Gordon models
via generalized zeta-function }
\author{Sergey Leble \\Gdansk University of Technology, \\Faculty of Applied Physics and Mathematics,
\\{\small ul. Narutowicza 11/12, 80-952
 Gdansk, Poland,}}
\maketitle

\begin{abstract}

One-dimensional Yang-Mills Equations are considered from a point
of view of a class of nonlinear Klein-Gordon-Fock models. The case
of self-dual Nahm equations and non-self-dual models are
discussed.  A quasiclassical quantization of the models is
performed by means of generalized zeta-function and its
representation in terms of a Green function diagonal for a heat
equation with the correspondent potential. It is used to evaluate
the functional integral and quantum corrections to mass in the
quasiclassical approximation.
  Quantum corrections to a few periodic (and kink) solutions of the Nahm as
  a particular case of the  Ginzburg-Landau
(phi-in-quadro) and   Sin-Gordon models are evaluated in arbitrary
dimensions. The Green function diagonal for heat equation with a
finite-gap  potential is constructed by universal description  via
solutions of Hermit equation. An alternative approach based on
Baker-Akhiezer functions for KP equation is proposed. The
generalized zeta-function is studied in both forms; its derivative
at zero point, expressed in terms of elliptic integrals is
proportional to  the quantum corrections to mass.
 \end{abstract}

\renewcommand{\abstractname}{\small }

\section{Introduction.}

\subsection{General remarks.}

We consider one-dimensional field theory, based on nonlinear
Klein-Gordon equations, arising, for example of Sine-Gordon (SG)
case, in kink models for crystal structure dislocations \cite{Br}
or in a context of a relativistic models \cite{Ket}. Popular
operator constructions of quantum field theory such as Yang-Mills
one have reductions to one dimensional models \cite{Nahm}. Some
kind of embedding of such theory into the multidimensional one is
possible: Atiyah-Drinfeld-Hitchin-Manin-Nahm construction may
appear as an equivalence between two sets of self-dual equations,
one as described above in one dimension, the other in three
dimensions (reduced from a Euclidean four dimensional theory by
deleting dependence on a single variable) \cite{Cor}.

A class of nonlinear Klein-Gordon equations in the case of static
one-dimensional solutions is reduced to
\begin{equation}
\label{KG}
\phi^{\prime\prime}- V^{\prime}(\phi)=0, \phi= \phi(x),
x\in R.
\end{equation}
Suppose the potential $V(\phi)$ is twice continuously
differentiable; it guarantees existence and uniqueness of the
equations correspondent to (\ref{KG}) Cauchy problem solution. The
first integral of the equation (\ref{KG}) is given by
\begin{equation}
\label{E}W = \frac{1}{2}\phi^{\prime2}-V(\phi),
\end{equation}
where W is the integration constant. The equation (\ref{E}) is
ordinary first-order differential equation with separated
variables. As the phase method shows, solutions of this equations
belong to the following families: constant, periodic, separatrix
and the so-called "passing" one \cite{TD}.

In the case of Nahm equation
\begin{equation}\label{N}
    V_{N}(\phi) =   \frac{\phi^{4}}{2}; \quad W =-\frac{w^4}{2},
\end{equation}
and for Ginzburg-Landau the "potential" is
\begin{equation}\label{GL}
    V_{gl}(\phi) = \frac{g}{4}(\phi^{2}-\frac{m^{2}}{g})^{2},
\end{equation}
while in the case of Sin-Gordon (SG) model it is
\begin{equation}
\label{Vsg}V_{sg}(\phi) = \frac{2m^{4}}{3g}(1+\\
cos(\frac{1}{m}\sqrt{\frac{3g}{2}}x)).
\end{equation}
We modified this last model to fit it with the first one at small
values of the constant g, namely
\[
V_{sg} = \frac{13m^{4}}{12g}+ V_{gl} + O(g^{2}).
\]

\subsection{One-dimensional reduction of Yang-Mills theory.}

Starting with the full Yang-Mills equations in four Euclidean
dimensions,
  \begin{equation}\label{YM}
   D_{\mu}T_{\mu\nu}=0,
\end{equation}
for the gauge fields $T_{\mu} = T_{\mu}^+$, where
$$
T_{\mu\nu} =T_{\nu,\mu} - T_{\mu,\nu}-\imath[T_{\mu},T{_\nu}],
\quad D_{\mu}\Phi = \partial_{\mu} - i [T_{\mu},\Phi].
$$
and demanding all fields be independent of three of the variables
$x_k$, k=1,2,3  setting $x_4 =z$,
\begin{equation}\label{YM1}
\begin{array}{c}
     \frac{d^2T_k}{dz^2} = [T_j[T_j,T_k]], \quad
  [T_k,\frac{d T_k}{dz}]=0. \\
\end{array}
\end{equation}
One cam easily check that the self-dual equations,
\begin{equation}\label{SDE}
    \frac{dT_i}{dz}=\pm \varepsilon_{ijk}T_jT_k,
\end{equation}
corresponding to the model,
 imply Eqs. (\ref{YM1}).
  Starting from the simplest solution of the system (\ref{SDE})
  \begin{equation}\label{ansatz}
    T_i = \phi_i\sigma_i
\end{equation}
   built on a base of Pauli matrices  one
arrives at the Euler system for $\phi_i(y)$ that is solved in
Jacobi functions. The solutions are dressed by the gauge-Darboux
transformations \cite{Leb}.

  For the second order equations the ansatz similar to (\ref{ansatz})
 \begin{equation}\label{ansatz1}
   T_i =\phi(z)\alpha_i , i=l,2,3,
\end{equation}
with a constant matrices $\alpha_i$  and a convenient choice of
scaling leads to the pair of equations
   \begin{equation}\label{nonSDE}
   \begin{array}{c}
       2 \alpha_i = \sum_{j=1}^{3}[\alpha_j[\alpha_j,\alpha_i]]      \\
   \phi''(z)=2\phi^3. \\
   \end{array}
 \end{equation}
 The second order equation for $\phi(z)$ enters as a particular case into
 the class  of nonlinear (\ref{KG}) with $V'(\phi)=2\phi^3$, or into Lagrangian of GL model  with
 V from (\ref{N}).

\subsection{Feynmann quantization of a classical field.}

An attention to Feynmann quantization formalism of a classical
field was recently attracted in connection with a link to a SUSY
quasiclassic quantization condition \cite{CC} suggested in
\cite{Ju}; see, however, \cite{Pe}.

Historically, quantum corrections started from a \cite{Dash}, see
also \cite{Raj}. Important development concerns the Jacobi variety
structure \cite{Smi}.

In the papers of V.Konoplich \cite{Ko} quantum corrections to a
few classical solutions by means of Riemann zeta-function are
calculated in dimensions $d > 1$. Most interesting of them are the
corrections to the kink - the separatrix solution of the field
$\phi^{4}$ (GL) model \cite{MB}.

The method of \cite{Ko} is rather complicated and it is desired to
simplify it, that was the main target of our previous note
\cite{LeZa4}. We applied the Darboux transformations technique
with some nontrivial details missed in \cite{Ko}.

The suggested approach open new possibilities; for example it
allows to calculate the quantum corrections to matrix models of
similar structure
, Q-balls \cite{CE} and periodic solutions of the models. The last
problem is posed in the review \cite{TD}.

The approximate quantum corrections to the solutions of the
equation (\ref{KG}) are obtained via the Feynmann functional
integral method evaluated by the stationary phase analog
\cite{Jun}. It gives the following relation
\begin{equation}
\label{S}\exp[-\frac{S_{qu}}{\hbar}]\simeq\frac{A}{\sqrt{det D}},
\end{equation}
where $S_{qu}$ denotes quantum action, corresponding the potential
$V(\phi)$, $A$ - some quantity determined by the vacuum state at
$V(\phi) = 0$, and $\det D$ is the determinant of the operator
\begin{equation}
\label{D}D = - \partial^{2}_{x} - \Delta_{y} +
V^{\prime\prime}(\phi(x)).
\end{equation}
The argument $y \in R^{d-1}$ stands for the transverse variables
on which the solution $\phi(x)$ does not depend. The operator D
appears while the second variational derivative of the quantum
action functional (which enter the Feynmann trajectory integral)
is evaluated. For the vacuum action $S_{vac}$ the relation of the
form (\ref{S}) is valid if $S_{qu}$ is changed to $S_{vac}$ and
$D$ is replaced by the "vacuum state" operator $D_{0} = -
\partial^{2}_{x} - \Delta_{y}+\nu$. Then, the quantum correction
\begin{equation}
\label{qucor}\Delta S_{qu} = S_{qu} - S_{vac},
\end{equation}
is obtained by the mentioned twice use of the formula (\ref{S}) as
\begin{equation}
\label{qucor1}\Delta S_{qu} = \frac{ \hbar}{2}\ln(\frac{\det{D}}{\det{D_{0}}%
}).
\end{equation}
Hence, the problem of determination of the quantum correction is
reduced to one of evaluation of the determinants $D$ and $D_{0}$
ratio. The methodic of the evaluation will be presented in the
following section.

\subsection{The generalized Riemann zeta-function and Green function of heat
equation.}

The generalized zeta-function appears in many problems of quantum mechanics
and quantum field theories which use the Lagrangian $\mathcal{L}%
=(\mathbf{\partial}\phi)^{2}/2  - V(\phi)$ and it is necessary to calculate a
Feynmann functional integral in the quasiclassical approximation.

The scheme is following. Let the set $\{\lambda_{n}\} = S$ be a
spectrum of a linear operator $L$,
 \begin{equation}
\label{lndet}\ln(\det L)= \sum_{\lambda_{n}\in S}\ln\lambda_{n},
\end{equation}
where the sum in the r.h.s. is formal one.

  The generalized
Riemann zeta-function $\zeta_{L}(s)$ is defined by the equality
\begin{equation}
\label{zet0}\zeta_{L}(s) = \sum_{\lambda_{n}\in
S}\lambda_{n}^{-s}.
\end{equation}

This definition should be interpreted as analytic continuation to
the complex plane of $s$ from the half plane $Re s > \sigma > 0 $
in which the sum in (\ref{zet0}) converges. Differentiating the
relation (\ref{zet0}) with respect to $s$ at the point $s=0$
yields
\begin{equation}
\label{lndet} \ln(\det L)= - \zeta_{L}^{\prime}(0).
\end{equation}
The generalized zeta-function (\ref{zeta}) admits the
representation via the diagonal $g_{L}$ of a Green function of the
operator $\partial_{t}+L$.

A link to the diagonal  Green function (heat kernel formalism) has
been used in quantum theory since works by Fock (\cite{F}).
There is a representation in terms of the formal sum over the
spectrum
\begin{equation}
\label{gL}g_{L}(t,\mathbf{{r},{r_{0}}) =\sum_{n} \exp[-\lambda_{n}
t] \psi
_{n}(r)\psi_{n}^{*}(r_{0}),}%
\end{equation}
where the normalized eigenfunctions $\psi_{n}(\mathbf{r)}$
correspond to eigenvalues $\lambda_{n}$ of the operator $L$.

Let us next define
\begin{equation} \label{defg}
 \begin{array}{c}
   \gamma_{L}\left(t\right)=\sum_{k}e^{-\lambda_{k} t}=\int g_{L}\left(t,\mathbf{{r},{r}}\right)d  \mathbf{r} \\
   \end{array}
 \end{equation}
The  Mellin transformation yields the generalized zeta function of
the operator L:
  \begin{equation}\label{gamma}
    \zeta_{L} (s)=\frac{1}{\Gamma(s)}
\int_{0}^{+\infty}t^{s-1}\gamma_{L}(t)dt.
\end{equation}
Returning to the operators (\ref{D})$ L\rightarrow D, D_0$:
  $D= - \Delta  +V^{\prime\prime}(\phi_{0}(x))$ and $D_{0}= -
 = - \frac{d^2}{dx^2}- \Delta_y+\nu$,
we pose
   the main problem for a periodic potential $u(x)$:
\begin{equation}\label{main}
   \left(\frac{\partial}{\partial t} + D_1\right) g_{D}\left(t,x,x_{0}\right) =\delta\left(t \right)\delta\left(x-x_{0}\right)
\end{equation}
where  $D_1=-\frac{d^2}{dx^2} + u(x)$, $V \leq V_m$, that means a
necessary presence of a continuous part of the spectrum $
 \lambda \in [\lambda_0, +\infty) $.
The basic relation (\ref{qucor1}) points to a necessity of
evaluation of the determinants of the operators
\begin{equation}
\label{det}D=D_{0}+u(x), \quad
D_{0}=-\partial_{x}^{2}-\Delta_{y}+\nu
\end{equation}
where $\lambda$ is a positive number and $x\in R$ is one of
variables, while $\mathbf{y}\in R_{d-1}$ is a set of other
variables. The operator $\Delta_{y} $ is the Laplace operator in
d-1 dimensions, u(x) is one-dimensional potential that is defined
by the condition
\begin{equation}
\label{pot}V^{\prime\prime}(\phi_{0}(x)) = \lambda+u(x),
\end{equation}
where $\phi_{0}(x)$ is the classical static solution of the
equation of motion.

The Green function for a Hermitian operator $D_1^+=D_1$ is :
$$g_{D_1}\left(t,x,x_{0}\right)=\sum_{k}e^{-\lambda_{k} t}\psi_k\left(x\right)\psi^{*}_k\left(x_{0}\right)\Theta\left(t \right).$$

The generalized zeta-function, defined by the relations (\ref{defg}%
,\ref{gamma}), will be referred as the zeta-function of the
operator $D$.

From the relation (\ref{defg}) for the function $\gamma_{D}(t)$ it
follows an important property of multiplicity: \textit{if the
operator D is a sum of two differential operators $D = D_{1} +
D_{2}$, which depend on different variables, the following
equality holds}
\begin{equation}
\label{gg}\gamma_{D}(t)=\gamma_{D_{1}}(t)\gamma_{D_{2}}(t).
\end{equation}

Generally, for a constant potential $\nu$ (see Sec. 3), for the
one-dimensional case,
\begin{equation}
\label{nu}\gamma_{D_{0}}(t) =   \frac{\exp [\nu t]}{2\sqrt{\pi
t})^d} .
\end{equation}
We will need the value of the function $\gamma_{D}(t)$ for the
vacuum state, when the operator $D= -\Delta_y$ is equal to the
d-1-dimensional Laplacian. In this case $\nu = 0$, hence
\begin{equation}
\label{Poi}\gamma_{D_{0}}(t) = \frac{1}{(2\pi)^{d-1}}\int_{\mathbb{R}^{d-1}}%
d\mathbf{k} \exp(-|\mathbf{k}|^{2}t) = (4\pi t)^{-(d-1)/2}.
\end{equation}
Combining as in  (\ref{gg}) yields
\begin{equation}
\label{nu}\gamma_{D_{0}}(t) =   \frac{\exp [\nu t]}{2\sqrt{\pi t}}
.
\end{equation}
 Then,
 \begin{equation}\label{zeta0}
    \zeta_{D_0}(s)=\frac{1}{\Gamma(s)}
\int_{0}^{+\infty}t^{s-1}\gamma_{D_0}(t)dt= \frac{\Gamma (s-\frac{1}{2}d)}{\Gamma (s)}\ \frac{1}{(2%
\sqrt{\pi })^{d}} \nu  ^{\frac{d}{2}-s\ }.
\end{equation}

A quantum correction to the action in one-loop approximation for
the classical solution $\phi(x)$ is calculated via zeta-function
by the formula
     \begin{equation}
\label{qucor2}\Delta S_{qu} = - \frac{ \hbar}{2}[\zeta^{\prime}_{D}(0) - \zeta^{\prime}_{D_0}(0)]/2 %
 .
\end{equation}

\section{The  classic static solutions and energy of solitons. }
\subsection{Static solutions of $\phi^4$. }
  In the case of static solutions of the $\varphi^4$ model the potential is determined by the formulas
\begin{equation}\label{Vphi4}
          V(\varphi) = \frac{g}{4}(\varphi^2- \frac{m^2}{g})^2,
\end{equation}
 therefore the
equation of motion has the form
\begin{equation}\label{eqmot}
    \varphi''(x) + m^2\varphi - g\varphi^3 = 0,
\end{equation}
that yields (\ref{E}) in the form
\begin{equation}\label{Wphi4}
(\varphi')^2 = \frac{g}{2}(\varphi^2- \frac{m^2}{g})^2 + 2W.
\end{equation}
Its restricted solutions, as it follows from phase plane analysis,
exist if
\begin{equation}\label{Wphi4cond}
   -\frac{m^2}{4g}\leq W \leq 0.
\end{equation}
The separatrix  (W=0) solution of (\ref{Wphi4}) is the kink/antikink
\begin{equation}\label{kink0}
    \varphi_0 = \pm\sqrt{\frac{2}{g}}b\tanh(bx)), \qquad b=\frac{m}{\sqrt{2}}.
\end{equation}
While inside the interval (\ref{Wphi4cond}) the equation
(\ref{Wphi4}) is expressed in terms of the elliptic Jacobi sinus
\begin{equation}\label{phi4sn}
   \varphi=\pm\sqrt{\frac{2}{g}}kb sn(bx;k), \qquad
   b=\frac{m}{\sqrt{1+k^2}}, \quad 0<k<1.
\end{equation}
The constant W is given by
\begin{equation}\label{Wsn}
    W = - (\frac{1-k^2}{1+k^2})^2\frac{m^4}{4g}.
\end{equation}
The energy in the Nahm case $k=i$ should be studied separately,
because the links (\ref{phi4sn},\ref{Wsn}) between $k,b$ and $W$
is not valid.

The family of the restricted solutions contains also the constant
vacuum ones (W=0)
\begin{equation}\label{phi4const}
    \varphi = \pm\frac{m}{\sqrt{g}}.
\end{equation}

After the substitution of (\ref{kink0}) into (\ref{pot}) we obtain
the following potential u(x) :
\begin{equation}\label{pokink}
    u(x) = -6b^2/ch^2(bx),
\end{equation}
with the meaning of the constant $  b = m/\sqrt{2}. $ As a result
the two-level reflectionless potential of one-dimensional
Schr{\accent "7F o}dinger equation $ -
\partial^2_x + u(x) $ appears. Eigenvalues  and the
normalized eigenfunctions of which are correspondingly (its
numeration is chosen from above to lowercase).
$$\lambda_1 = - b^2, \qquad \psi_1(x) = \sqrt{3b/2} \sinh(bx)/cosh^2(bx);$$
$$\lambda_2 = - 4b^2,\qquad \psi_2(x) = \sqrt{3b}/2\cosh(bx).$$

In the case of  the periodic solution of the phi-in-quadro model
the potential has a "cnoidal" form
\begin{equation}\label{cnphi4}
   u(x)= - 6k^2b^2cn^2(bx;k)+(5k^2 - 1)b^2.
\end{equation}
This potential differs from second Lam$\acute{e}$s equation
 potential by the constant hence its spectrum is two-gap one.
 \subsection{SG  model}
Let us briefly describe SG model \cite{LeZa0} we integrate the
equation (\ref{KG}) with
 the potential (\ref{Vsg}) arriving at the first-order differential
 equation
 with the parameter W (\ref{E}).
 \begin{equation}\label{Wsg}
   (\varphi')^2 =
   \frac{4m^4}{3g}(1+cos(\frac{1}{m}\sqrt{\frac{3g}{2}}\varphi))+
   2W.
 \end{equation}
The solutions are restricted and hence have the direct physical
relevance, if
\begin{equation}\label{condW}
    -\frac{4m^4}{3g}\leq W \leq 0,
\end{equation}
that follows from phase plane analysis. If $W=0$ the nontrivial
solutions are interpreted as kink and antikink
\begin{equation}\label{kink}
    \varphi(x)=\pm\sqrt{\frac{2}{3g}}\arcsin{\tanh(mx)}(mod\Phi),\qquad
    \Phi=2m\pi \sqrt{\frac{2}{3g}},
\end{equation}
while at the interval
$$
  -\frac{4m^4}{3g}< W < 0,
$$
the solution of (\ref{Wsg}) yields a periodic function expressed via
elliptic Jacobi function. To find it one plug $\varphi = \pm 2m
\sqrt{\frac{2}{3g}}\arcsin{z},$ then the equation (\ref{Wsg}) goes
to
\begin{equation}\label{sJac}
    \left(z'\right)^2=m^2\left(1+\frac{3Wg}{4m^4}-z^2\right)\left(1-z^2\right)
\end{equation}

The solution of (\ref{sJac}) at the interval (\ref{condW}) is
given by
\begin{equation}\label{z=sn}
    z=ksn(mx;k),
\end{equation}
where
\begin{equation}\label{k}
    k=\sqrt{1+\frac{3gW}{4m^4}}
\end{equation}
is the module of the elliptic function. Hence

\begin{equation}\label{W}
    W = \frac{4(k^2-1)m^4}{3g}.
\end{equation}
Finally
\begin{equation}\label{varphi}
   \varphi = \pm
2m \sqrt{\frac{2}{3g}}\arcsin{ksn(mx;k)}\quad (mod\Phi).
\end{equation}
The class of restricted solutions contains also
\begin{equation}\label{0}
    \varphi=0 (mod\Phi); W=-\frac{3m^4}{3g}
\end{equation}
and
\begin{equation}\label{const}
    \varphi=\pm\pi m \sqrt{\frac{2}{3g}} (mod \Phi) ; W=0.
\end{equation}
Other static solutions are obtained by shifts
$$
x \rightarrow x+x_0, \quad \varphi\rightarrow \varphi + \Phi,
$$
that follows from Klein-Gordon equation invariance and SG equation
potential periodicity.

Let us evaluate the energy of the nontrivial static solutions of
both models via the  energy density definitions
\begin{equation}\label{Ekink}
e(x) = \int_{-\infty}^{\infty}((\phi')^2/2+V(\varphi))dx,
\end{equation}
for kinks and, in a case of periodic solutions,
\begin{equation}\label{Ekink}
E = 2\int_{0}^{l}e(x)dx,
\end{equation}
 the constant "l" is the period of
the solution.

For the SG kink:
\begin{equation}\label{Ekink}
    E_k = \frac{16m^2}{g},
\end{equation}
and, for a periodic soliton
\begin{equation}\label{Eper}
     E_p = \frac{8m^2}{g}[(1-k^2)K+2E],
\end{equation}
where K(k),E(k) - complete elliptic Legendre integrals.

 \subsection{Static solutions of  Nahm model as a specific case of $\varphi^4$ model}

The solution of the (\ref{nonSDE}), is a particular but specific
case of static GL model (m=0, g=2)
\begin{equation}
\phi^{\prime2}=\phi^{4}-w^{4}.\label{Nahmphi4}%
\end{equation}
A solution of (\ref{nonSDE}) is expressed via elliptic functions
\cite{Erd}, namely while
\begin{equation}
\int_{0}^{\phi}\frac{d\phi}{\sqrt{\phi^{4}-w^{4}}}=\int_{0}^{\phi}\frac{d\phi
}{\sqrt{\left(  \phi^{2}-w^{2}\right)  \left(  \phi^{2}+w^{2}\right)  }%
}=z\ ,\label{dphi}%
\end{equation}
or
\bigskip $\frac{1}{w}\int_{0}%
^{\frac{\phi}{w}\ }\frac{dt}{\sqrt{\left(  t^{2}-1\right)  \left(
t^{2}+1\right)  }}=\frac{1}{w}\int_{0}^{\frac{\phi}{w}\ }\frac{dt}%
{i\sqrt{\left(  1-t^{2}\right)  \left(  1-i^{2}t^{2}\right)  }}.$

Hence the solution is
\begin{equation}
\phi=wsn(iwz,i),\label{wsn}%
\end{equation}
 that may be tested by direct differentiation  in (\ref{Nahmphi4}).

The invariants
\begin{equation}\label{invN}
    \begin{array}{c}
       g_2=w^4, \\
       g_3=0, \\
      \end{array}
\end{equation}
alternatively determine the potential in terms of the Weierstrass
function,
\begin{equation}\label{phiNahm}
   \phi(z,X)=w+\frac{w^3}{\wp(z;w^4,0)-w^2/2}.
\end{equation}
or, if take into account he relation
\begin{equation}\label{WJ}
    \wp(z;w^4,0)=e_3+\frac{e_1-e_3}{sn^2(\sigma z)} , \quad \sigma
=\sqrt{e_1-e_3},\quad k=\sqrt{ \frac{e_2-e_3}{e_1-e_3}}
\end{equation}
one has for $e_3=0, e_1=-e_2=w^2/2,$ the parameters of the
solutions in Jacobi terms $k=i,\sigma=w/\sqrt{2} $
\begin{equation}\label{nSDEsol}
  \phi(z)=\sigma \frac{sn[\sigma(z-z_0)] dn[\sigma(z-z_0)] }{cn[\sigma(z-z_0)]}.
\end{equation}
This form coincideds with one from \cite{Corr}. We will use the
expression of the potential in terms of the solution (\ref{wsn})
 \begin{equation}\label{potnam}
\bigskip V^{\prime\prime}(\phi_{0}(x))=\left(  \frac{\phi^{4}}{2}\right)
^{\prime\prime}=6\phi^{2}=-6\sigma^{2}\left( 1-cn^{2}(\sigma
z,i)\right).
\end{equation}

\section{The generalized zeta-function via Hermit equation}

In a spirit of Hermit approach, see, e.g. \cite{Mos} , the
function $\hat{g}_L(p,x,x) = \int \exp[-pt] g_L(t,x,x) dt$ is a
solution of bilinear equation
\begin{equation}\label{Hermit}
    2GG'' - (G')^2 - 4(u(x)+p)G^2+1=0,\qquad G(p,x) = \hat{g}_L(p,x,x)
\end{equation}
which in a case of reflectionless and finite-gap solutions is
solved more effectively than (\ref{L+p}). It is possible to cover
all necessary classes of solutions of the models ( A,B for SG, C,D
for $\phi^4$) case and  $D_0$ for Nahm, via the universal
representation by means of polynomials  (in p) $P,Q$
\begin{equation}\label{PQ}
    G(p,x)=P(p,z)/2\sqrt{Q(p)},
\end{equation}
where
$$z = sech^2(bx)$$
for kinks A,C, and
$$
z=cn^2(bx;k)
$$
for the periodic B,D.
\begin{equation}\label{linkPQ}
    b^2(\rho(z)(2PP''-(P')^2)+\rho'(z)PP')-(p+u(z))P^2+Q=0,
\end{equation}
the primes denote derivatives with respect to z, while
\begin{eqnarray} \label{pots}
  \rho &=& \{\begin{array}{c}
     \\        z^2(1-z), cases \quad A,C; \\
               z(1-z)(1-k^2+k^2z), cases \quad D_0, B,D. \\
             \end{array}\\
  u(z) &=& \{\begin{array}{c}
               b^2(1-2z), case \quad (A) \\
       b^2(2k^2-1-2k^2z), case \quad (B) \\
                     2b^2(2-3z), case \quad (C) \\
              b^2(5k^2-1-6k^2z), case \quad (D) \\
 -6b^{2}\left( 1-z\right)         case         \quad (D_0)\\
                  \end{array}
\end{eqnarray}

Substituting (\ref{RQB}) into (\ref{linkPQ}) gives for each power
of $p = 0,1,2$
\begin{equation}\label{sys}
    \begin{array}{c}
       -2P_1(z)-u(z)+q_2=0, \\
       b^2(2\rho(z)P_1''+\rho'(z)P_1')-P_1^2 - 2u(z)P_1+q_1=0.\\
        b^2(\rho(z)(2P_1P_1''-P_1'^2)+\rho'(z)P_1P_1'-u(z)P_1^2+q_0=0. \\
     \end{array}
\end{equation}
respectively.

Let us start with the cases (A,B). The form of the polynomial is
determined from well-known facts of the reflectionless potentials
theory. The substitution
\begin{equation}\label{RQB}
    P=p+P_1(z), \quad Q=p^3+q_2p^2+q_1p+q_0
\end{equation}
into (\ref{sys}) results in
\begin{equation}\label{P1}
    P_1=k^2b^2z, \quad    q_2=b^2(2k^2-1),\quad q_1=b^4k^2(k^2-1), \quad q_0=0
\end{equation}
  includes (A) in a sense that for the
case  $k=1$  (\ref{P1}) gives
\begin{equation}\label{Arez}
    P_1=b^2z.
\end{equation}

Going to the cases (C,D), generally
\begin{equation}\label{PQD}
    P=p^2+P_1(z)p+P_2(z), \quad
    Q=p^5+q_4p^4+q_3p^3+q_2p^2+q_1p+q_0,
\end{equation}
A substitution of (\ref{PQD}) into the Hermit equation splits in the
system
\begin{equation}\label{splitD}
    \begin{array}{c}
       -2P_1  - u  + q_4 = 0, \\
    -2P_2  - P_1^2  - 2 u P_1 +b^2(2\rho P_1''+\rho'P_1')+q_3=0,    \\
       b^2(\rho (2P_2''+2P_1P_1''-(P_1')^2)+\rho'(P_2'+P_1P_1'))-2P_1P_2)-u(2P_2+P_1^2)+q_2 =0 ,\\
       b^2(2\rho(2P_1''P_2-P_1'P_2'+P_1P_2'')+\rho'(P_1P_2'+P_1'P_2))-P_2^2 - 2uP_1P_2 + q_1=0,\\
       b^2(\rho(2P_2P_2''-P_2'^2)+\rho'P_2P_2')-uP_2^2+q_0=0.\\
     \end{array}
   \end{equation}
The potentials  $u(z)=c-6b^2k^2z$, yields for the case D
\begin{equation}\label{P1P2}
P_1(z)=\alpha z + \beta, \qquad P_2(z)=a'z^2 +b' z+c',
\end{equation}
where
\begin{equation}\label{alpha}
    \begin{array}{c}
      \alpha = 3k^2b^2, \\
      \beta =3b^2,\\
      a' = 18b^4k^4, \\
      b' =  -3b^2k^2(11b^2k^2-q_4-b^2),  \\
      c' =63b^4k^4/4-5b^2k^2q_4/2+b^2q_4/2- q_4^2/4+q_3- 9b^4k^2/2+3b^4/4,\\
    \end{array}
\end{equation}
are functions only of $k,b$, where
\begin{equation}\label{q}
    \begin{array}{c}
       q_0 = 0, q_1= -27k^2(1-k^2)^2b^8, \\
        q_2=-9b^6(k^2+1)(k^4-4k^2+1), q_3 = 3b^4(1+9k^2+k^4), q_4=5b^2(1+k^2).  \\
      \end{array}
\end{equation}
 Finally,
 \begin{equation}\label{Q}
    Q=\prod_{i=1}^{i=5}(p-p_i),
\end{equation}
where the polynomial $Q$ simple roots $p_i$ are ordered so that
($0<k<1$)
\begin{equation}\label{p_i}
    - (2\sqrt{1-k^2+k^4}+1+k^2)b^2< -3b^2<-3k^2b^2<0<-
    (2\sqrt{1-k^2+k^4}-1-k^2)b^2.
\end{equation}
For the cases $(D_0, C,D)$ we begin from   that it is the result of
substitution of $P_1$ from the first equation of (\ref{sys}) into
the second one. Next,  the third equation yields
\begin{equation}\label{Qrez}
    Q=p(p^2+(1-k^2)b^2)(p-k^2b^2),
\end{equation}
with the simple roots $p_i$. while in the particular case of
$\phi^4$ (C)
$$
q_0=0,\quad q_1=0,q_2=36b^6,\quad q_3=33b^4,q_5=10b^2.
$$
The arguments in (\ref{splitD}) are omitted.

Let us pick up the expressions determining $\hat{\gamma}(p)$:
\begin{equation}\label{hatgammap}
    \hat{\gamma}(p) = \int\frac{P(z)}{2\sqrt{Q}}dx =
    \frac{p^2}{2\sqrt{Q}}\int dx+ \frac{p}{2\sqrt{Q}} \int(\alpha z+\beta) dx +
    \frac{1}{2\sqrt{Q}} \int(a'z^2 +b' z+c') dx.
\end{equation}

The case of Nahm equation   yields
  $q_{4}=0,\quad q_{3}=\allowbreak
-21b^{4},q_{2}=\allowbreak0, \,
q_{1}=\,108b^{8},q_{0}=\allowbreak0, $ hence $P_1(z)=
-3b^2(z-1),P_2=18b^4z^2-36b^4z.$
  \begin{equation}\label{Q}
    Q(p)=\prod_{i=1}^{i=5}(p-p_i) =p(p+3b^2)(-p+3b^2)(2\sqrt{3}b^2-p )(2\sqrt{3}b^2+p ),
\end{equation}
where the polynomial $Q$  simple roots $p_i$ are easily ordered
for real $b$.

So we need three integrals over the period $K$.
\begin{equation}\label{intz}
    \begin{array}{c}
       \int_0^K dx ,
       \int_0^K z dx  ,
\int_0^K z^2dx.
           \end{array}
\end{equation}
Let us go to the variable z, $ dz=d(cn^2(x))= - 2cn(x)sn(x)dn(x)dx
 $, a bit more convenient to put $y=1-z, dy=-dz$
\begin{equation}\label{inty}
     \begin{array}{c}
        \int_0^K dx = \int_0^1\frac{dy}{\overleftarrow{y(1-y)(1-k^2y)}} = 2K(k), \\
       \int_0^K sn^2(x;k)dx = \frac{1}2{}\int_0^1\frac{ydy}{\sqrt{y(1-y)(1-k^2y)}} = \frac{K(k)-E(k)}{k^2}, \\
       \int_0^K sn^4(x;k) dx = \int_0^1\frac{y^2dy}{\sqrt{y(1-y)(1-k^2y})} = \frac{1}{3k^4}( (2+k^2)K(k)  - 2 (1+k^2)E(k)) . \\
     \end{array}
\end{equation}
hence
\begin{equation}\label{hatgammap'}
\begin{array}{c}
  \hat{\gamma}(p) = 2K(k)\frac{p^2}{2\sqrt{Q}} + \alpha (-\frac{K(k)-E(k)}{k^2}+ (\beta -\alpha)2K(k)\frac{p}{2\sqrt{Q}}+ \\
  (a' \frac{(2+k^2)K(k)  - 2 (1+k^2)E(k)}{3k^4} + (b'-2a') \frac{K(k)-E(k)}{k^2}+
    (a' -b' +c')2K(k)) \frac{1}{2\sqrt{Q}}.\\
\end{array}
   \end{equation}
or, finally
\begin{equation}\label{zeta}
    \zeta(s) =  \int_0^\infty (\int_l \hat{ \gamma}(p)e^{pt}dp)t^{s-1}dt/\Gamma(s);
\end{equation}
which is expressed via integrals
\begin{equation}\label{intp}
  \begin{array}{c}
                                     \int_l \frac{e^{pt}}{\sqrt{Q}} dp \\
                                      \int_l \frac{pe^{pt} }{\sqrt{Q}} dp  \\
                                     \int_l \frac{p^2e^{pt}}{\sqrt{Q}} dp \\
                                  \end{array}
\end{equation}
by a contour that contains all branch points of the integrands
(inverse Laplace transform). Or, via
\begin{equation}\label{beta}
     \int_0^\infty  e^{pt}t^{s-1}dt= \int_0^\infty  e^{t}(\frac{t}{p})^{s-1}d\frac{t}{p}=-\frac{1}{(-p)^{s-2}}\int_0^\infty  e^{-t}t^{s-1}dt=-\frac{1}{(-p)^{s}}\Gamma(s),
\end{equation}
$Im p<0$,  one arrives at
\begin{equation}\label{zeta}
    \zeta(s) = - \int_l \frac{\hat{ \gamma}(p)}{(-p)^{s}} dp.
\end{equation}
In the case of Nahm the mass depends on b:
\begin{equation}
\zeta (s)=-\ \int_{l}\frac{1}{(-p)^{s}}\frac{%
2K(i)p^{2}+3b^{2}(K(i)-E(i))p-48b^{4}K\left( i\right) }{2\sqrt{%
p(p+3b^{2})(-p+3b^{2})(2\sqrt{3}b^{2}-p)(2\sqrt{3}b^{2}+p)}}dp.
\label{zeta}
\end{equation}%
The result is given by hyperelliptic integral that can be
evaluated numerically or via elliptic functions if the symmetry of
integrand is taken into account.

\section{The generalized zeta-function as a combination of Baker-Achiezer functions.}

\medskip

Let us consider the  Green
function defined in the Sec.1 by the relation 
(\ref{gL}). The method we use here is based on the technique of
finite-gap integration of the KP equation \cite{IBBEM,M,Dub}.

The source part of the problem is one-dimensional version of the
heat equation (\ref{main})  with the  potential   $u(x) =
V''(\phi(x))$
\begin{equation}\label{g2}
    \left(\frac{\partial}{\partial t} - \frac{d^2}{dx^2}+ u(x)\right) G\left(t,x,x_{0} \right)=\delta\left(t
\right)\delta\left(x-x_{0}\right).
\end{equation}

 Let us integrate the equation (\ref{g2}) by x over $[x_0-\epsilon,
x_0+\epsilon]$, hence
\begin{equation}\label{L+p}
\begin{array}{c}
  \lim_{\epsilon \rightarrow 0}  \int_{x_0-\epsilon,
}^{x_0+\epsilon}\left(\frac{\partial}{\partial t} -
\frac{d^2}{dx^2}+u(x)\right)G(t,x,x_0)dx = \\
   \lim_{\epsilon \rightarrow 0}[-\frac{dG(t,x_0+\epsilon,x_0,x_0)}{dx} +
\frac{dG(t,x_0-\epsilon,x_0)}{dx}] =  \lim_{\epsilon \rightarrow
0}\int_{x_0-\epsilon, }^{x_0+\epsilon}\delta\left(t \right) \delta(x-x_0)dx = \delta\left(t \right).   \\
\end{array}
\end{equation}
The function $G(t,x,x_0)$ is supposed to be continuous at $t>0$.

 Solutions of the equation (\ref{g2}) with the zero r.h.s. are
expressed in terms of the Baker-Achiezer functions $\psi_s(x,t;P)$
built by the polynomial $q(s)=sx+s^2t$ of local parameter s
\cite{IBBEM,Dub}. Consider a Riemann surface of genus g and a
theta function on it. Then there are holomorphic differentials
$\omega_k, \quad k=1,...,g$, normalized as
\begin{equation}\label{omeganorm}
   \int_{a_i} \omega_k  =2\pi \delta_{ik}.
\end{equation}
  The non-special poles divisor on the surface is denoted as $D=P_1+...+P_g$ and the matrix of periods
is
\begin{equation}\label{Bik}
    B_{jk}=\int_{b_j}\omega_k,
\end{equation}
that define the general Riemann Theta:
 \begin{equation}\label{Rteta}
     \Theta (z)= \Theta(z|B) = \sum_{N \in Z_g} \exp[(BN\cdot N)/2+(N\cdot z].
\end{equation}
The parametrization by the polynomial $q(s)$ yields
\begin{equation}\label{psiBA}
   \psi_s(x,t;P) =C(P)\exp(\int_{P_{\infty}}^P[d\Omega^{(1)}x+d\Omega^{(2)}t] )
    \frac{\Theta(xU+tV+A(P) +D)}{\Theta(xU+tV+ D )},
\end{equation}
where $A(P)=\int_{P_{\infty}}^P\omega$ is Abelian map, $C(P)$ is a
constant and the vectors of b-periods
\begin{equation}\label{UV}
    U_i=\int_{b_i}d\Omega^{(1)}, \quad
V_i=\int_{b_i}d\Omega^{(2)}
\end{equation}
define the argument of the $\Theta$-function.
 The link between the
potential $u(x)$ and $\psi$ is given by
\begin{equation}\label{upsi}
    u(x)=  - 2\frac{\partial^2}{\partial x^2}  \ln  \Theta(xU+tV+D)
\end{equation}
which is recognized as general Matveev-Its formula \cite{IBBEM}.

So the parameters of the potential and the Green function arise
from (\ref{UV}) as the position of the cycles $b_i$ depends on
parameters of the potential via the curve that determine the
Riemann surface.

 Applying the classical
method, one builds the Green function which has the form prescribed
by (\ref{L+p}) via the linear independent solution $ \psi_{-
s}(x,t;P)$ as:
\begin{equation}\label{Grx}
    G_0(t,x,x_0)=\frac{1}{M}\psi_{s}(x,t;P) \psi_{- s}(x_0,t;P)
\end{equation}
that account the unit jumps of the first derivative with respect
to x at  $x=x_0$ and the Wronskian
$M=\psi_{s}\psi_{-s}'-\psi_{-s}\psi_{s}'$ is determined by a
normalization at the borders of the period.

The boundary condition at $t=0$ is satisfied if integrate
(\ref{Grx}) with respect to s along a contour C determined by the
spectrum of the operator $D_1$
\begin{equation}\label{dkBA}
   \int_C ds \frac{1}{M}\psi_{s}(x,t;P) \psi_{-s}(x_0,t;P).
\end{equation}
The zero value  at $t<0$  is guaranteed by the theorem on Laplace
transform \cite{Erd}.

 The diagonal values of the Green function are given by
\begin{equation}\label{Gryd}
\begin{array}{c}
   G(t,x,x) = \int_C \frac{ds }{M}\psi_{s}(x,t;P) \psi_{-s}(x,t;P) =  \\
  \int ds
\exp(\int_{P_{\infty}}^Pd\Omega+\int_{P_{\infty}}^{P_{-}}d\Omega))
    \frac{\Theta(xU+tV+A(P) +D)}{\Theta(xU+tV+D ) }
    \frac{\Theta(xU+tV+A(
P_-)+D)}{\Theta(xU+tV+D))  }  \\
\end{array}
\end{equation}
$P_-$ corresponds to $q_{-s}=- sx+s^2t$. The function $\gamma(t)$
and, next, the zeta-function are immediately written as
(\ref{gamma}).

The case of SG illustrates the idea in the simplest manner. We
start from the Lame equation,
 \begin{equation}\label{Lame}
 \bigskip\left( -\frac{d^{2}}{du^{2}}+2\bigskip\wp(u)\right)
\Psi=H\Psi,
\end{equation}
which solutions
\begin{equation}
\Psi=\frac{\sigma(u+\wp^{-1}(-H))}{\sigma(u)}\exp[\varsigma(\wp^{-1}(-H))u],
\label{psiL}%
\end{equation}
form a complete set (see e.g. \cite{Dub,BL}). The linear independent solution is obtained by %
the change of the sign before  $\wp^{-1}(-H))$.

\bigskip The equation (\ref{Lame}) is directly connected with one
with the cnoidal potential (\ref{pot})
\begin{equation}
\left(  -\frac{d^{2}}{dz^{2}}+2k^{2}-2k^{2}cn(z,k)^{2}\right)
\Psi=h\Psi,
\label{LamJ}%
\end{equation}
where $z=iK^\prime+u\sqrt{e_{1}-e_{3}}$, $H=\left(
e_{1}-e_{3}\right)
h+2e_{3}, k^{2}=\frac{e_{2}-e_{3}}{e_{1}-e_{3}},e_{1}=(2-k^{2}),\ e_{2}%
=(2k^{2}-1),e_{3}=-(1+k^{2}),$
 and v.v.
$e_{i}=\wp(\omega_{i}),\omega_{1}=\omega,\omega_{2}=\omega-\omega^{\prime
},\omega_{3}=\omega^{\prime}.$ The link between Weierstrass and
Jacobi functions
\begin{equation}\label{W-J}
    sn^2(w,k)=\frac{e_{1}-e_{3}}{\wp(z)-e_3},\quad
w=z\sqrt{e_{1}-e_{3}},
\end{equation}
 is obviously used.
It allows to express a solution of (\ref{LamJ}) as%

\begin{equation}
\Psi=\frac{\sigma(\frac{z-iK^\prime}{\sqrt{e_{1}-e_{3}}}+\wp^{-1}(\left(
e_{1}-e_{3}\right) h+2e_{3}))}{ \sigma(\frac{z-iK^\prime}{\sqrt
{e_{1}-e_{3}}})}\exp[\frac{z-iK^\prime}{\sqrt{e_{1}-e_{3}}}\varsigma(\wp^{-1}(\left(
e_{1}-e_{3}\right)
h+2e_{3}))],\label{psiJ}%
\end{equation}
 $e_{1}-e_{3}=3,\wp^{-1}(H)=\rho,h=\frac{\wp(\rho)-2e_{3}}{e_{1}-e_{3}%
};u=\frac{z-iK^\prime}{\sqrt{e_{1}-e_{3}}} \quad K'(k')%
$ is the complete elliptic integral.   The parameters are
expressed in terms of half-periods of the Weierstrass function,
defined by the curve with the uniformization  $\left(
\wp^{\prime}(u)\right) ^{2}=4\left( \wp(u)\right)
^{3}-g_{2}\wp(u)-g_{3}$.

It is convenient to perform numerical calculations using Jacobi
theta-function , namely
 $\sigma(u)=2\omega\exp[\frac{\eta
u^{2}}{2\omega}]\frac{\vartheta\left( \frac{u}{2\omega}\right)
}{\vartheta^{\prime}\left(  0\right)  }, \quad$
$\eta=\varsigma(\omega),$ that leads to
\begin{equation}
\Psi=\frac{\exp[\frac{\eta(2u\rho+\rho^{2})}{2\omega}]\vartheta\left(
\frac{u+\rho}{2\omega}\right)  }{\vartheta\left(  \frac{u\
}{2\omega}\right)
}.%
\end{equation}
The particular elliptic solution of the heat equation with the
potential as in (\ref{LamJ}) is the product
$\psi(z,t,h)=\exp[-ht]\Psi(z)$.

\section{BA function and "twisted BPS monopoles"}

 The  BA function (Its-Matveev formula)  for SG case
 is expressed in terms of Jacobi theta functions and solves the
 correspondent
Lame equation with $u_L(x)=2k^2(1-cn^2(x,k))$ and the spectral
parameter h \cite{Erd}
\begin{equation}\label{psi+}
\psi_{+}(x;h)=\frac{\exp[\frac{\eta(2v\rho+\rho^{2})}{2\omega}+v\zeta(\wp
^{-1}(3h\ -2(1+k^{2})))]\vartheta\left(
\frac{v+\rho}{2\omega}\right)
}{\vartheta\left(  \frac{v\ }{2\omega}\right)  },%
\end{equation}
where $\rho=\ \wp^{-1}(3h -
2(1+k^{2})),\eta=\zeta(\omega),\omega=\wp^{-1}(2-k^{2}),v = (x -
iK')/\sqrt{3}, \quad K'(k')%
$ is the complete elliptic integral, $\wp$ is the Weierstrass
function, $\zeta(\omega)$ is the Weierstrass zeta
 function ($\zeta'=-\wp$) \cite{Erd}.%
The potential $u_L(x)$ of the standard Lam$\acute{e}$ equation
differs from one $u(x)$ by the constant factor and shift, namely
$u_L=b^2u+b^2$, compare with (\ref{pots}). In this section we
choose for simplicity of formulas $b=1$.  Hence the spectral
parameters are connected by $h=p+1$.

 The basic theta function of the representation (\ref{psi+}) is defined by
\begin{equation}\label{teta}
    \vartheta\left( w  \right)= \vartheta\left( w |\tau \right) =
    \sum_{m=-\infty}^{\infty}\exp[i\pi(m^2\tau+2mw)],
\end{equation}
where $exp[i\pi \tau]=\eta$, $Im \,\tau$ should be positive for
the series convergence. The series convergence is rapid, therefore
the representation (\ref{psi+}) is convenient for numeric
evaluation of the integrals in the zeta formalism.

 The Green function $g_h$ of the spectral Lam$\acute{e}$ problem may be constructed as a product of two
independent solutions $\psi_+, \psi_-$ of the spectral equation
with the same h:
\begin{equation}\label{g_h}
    g_h(x,x_0)=\frac{1}{\mathbb{W}} \{\begin{array}{c}
                                            \psi_{+}(x;h)
  \psi_{-}(x_0;h),\quad x<x_0\\
        \psi_{-}(x;h)
  \psi_{+}(x_0;h)                                   \quad x>x_0
                                         \end{array}
   .
\end{equation}
The Wronskian factor $\mathbb{W}$ is chosen to normalize
(\ref{Gryd}) so that account the unit jumps of the first
derivative with respect to x:
\begin{equation}\label{g_h}
    \lim_{\epsilon \rightarrow 0}[\frac{dg_h(x_0+\epsilon,x_0,x_0)}{dx}
    -
\frac{dg_h(x_0-\epsilon,x_0)}{dx}]=-1.
\end{equation}
The independent solution $\psi_{-}(x,t;h) $ may be chosen
antisymmetric with respect to the reflection $x \rightarrow -x$,
e.g defined via $\vartheta(w+\frac{1}{2}).$ .

The boundary condition at $t=0$ may be account via the integral by
spectrum
 (see also (\ref{gL})) of the linear independent solutions product
\begin{equation}\label{Gryd}
  g(t,x,x_0) = \int\frac{1}{\mathbb{W}}\{\begin{array}{c}
                                           \Psi_{+}(x,t;h),
  \Psi_{-}(x_0,t;h), \quad x<x_0 \\
                      \Psi_{+}(x_0,t;h),
  \Psi_{-}(x,t;h),
                     \quad x>x_0                    \end{array}\}
  dh.
\end{equation}
where $\Psi_{+}(x,t;h)= \exp[-ht]\psi_{+}(x;h)$.

 Finally we integrate the diagonal values of the Green function $\gamma_{D_1}(t,x)=\int g(t,x,x) $ from (\ref{Gryd})
by the period of a solution
 and, after that, one can use the definition (\ref{zeta}) of the
generalized zeta-function via (\ref{Gryd}). Integrals dependence
on $z_i$ are expressed via the Weierstrass function of w. It is
also known, that
\begin{equation}\label{tetaper}
    \theta(z+B)=\exp[-\frac{B}{2}-z]\theta(z),
\end{equation}
so the ratio of the theta-functions has the period B.

\section{SG kinks}

The results of the previous sections allow to evaluate corrections to actions
for all four (A,B,C,D) cases. The results for $\phi^{2}$ model kinks are
well-known, see, e.g. \cite{LeZa4}, where a table for the dimensions d=1,2,3,4
is listed. These results fit the case B, which hence was strictly verified.

Let us present the formulas for the kinks of the SG model. The substitution of
the expressions from (\ref{PQ},\ref{Q},\ref{RQB},\ref{Arez}) yields a
divergence of the integral for (\ref{hatgammap}) $\hat{\gamma}(p)$. To
regularize the Green function let us divide itd Laplace transform into two
parts as
\begin{equation}
\label{Gck}G(p,x) = G_{c}(p) + G_{k}(p,x),
\end{equation}
where the part $G_{c}(p)$ is the Green function diagonal (the solution of
(\ref{Hermit}) for a constant potential:
\begin{equation}
\label{Gc}G_{c}=\frac{1}{2\sqrt{p+b^{2}}}.
\end{equation}
The kink part is easily constructed via (\ref{PQ},\ref{Q},\ref{RQB}%
,\ref{Arez}):
\begin{equation}
\label{Gk}G_{k}=\frac{b^{2}sech^{2}(bx)}{2p\sqrt{p+b^{2}}}.
\end{equation}
The representation (\ref{Gck}) results in two contributions of the quantum
corrections to the kink (antikink) mass.

The first one coincides with (\ref{nu}) for $\nu=m^{2}$.
\begin{equation}
\label{nu}\gamma_{D_{0}}(t) = (4\pi)^{\frac{1-d}{2}}\frac{\Gamma(s+\frac
{1-d}{2})}{\Gamma(s)}m^{ d-1-2s },
\end{equation}
and the second one is obtained by the general scheme, namely
\begin{equation}
\label{gammak}\gamma_{k}(p) = \int_{-\infty}^{\infty}G_{k}(p,x)dx.
\end{equation}
Plugging (\ref{Gk}) into (\ref{gammak}) yields
\begin{equation}
\label{gamkp}\gamma_{k}(p)=\frac{b}{p\sqrt{p+k^{2}}}.
\end{equation}
The value of the integral $\int_{-\infty}^{\infty}sech^{2}(bx) = 2/b$ is taken
into account. the corresponding function (\ref{gamma}) or, more exactly
(\ref{gammat}) will be denoted $\gamma_{L}^{R}$, the index R label the result
of a renormalization provided by the division (\ref{Gck}). The function may be
found directly from a table (\cite{BE}) but we would explain the result as an
example for further development of the renormalization procedure. Note that
the cut for the radical $\sqrt{p+b^{2}}$ is made along the Re p -axis from $-
\infty$ to $-b^{2}$, the branch with $\sqrt{p+b^{2}} = i\sqrt{|p+b^{2}|}$ on
the upper and $\sqrt{p+b^{2}}=-i\sqrt{|p+b^{2}|}$ on the lower bounds of the
cut is chosen. After the deformation of the integral contour one has for $t >
0$
\begin{equation}
\label{gamRt}\gamma_{L}^{R} = 1 - \frac{b}{\pi}\int_{0}^{\infty}\frac{\exp
(\xi+ b^{2})t}{(\xi+b^{2})\sqrt{\xi}}d \xi.
\end{equation}
The formula (\ref{gamRt}) gives an expansion of the resolvent of the operator
$\partial_{t}+L$ by the operator $L$ spectrum. The direct substitution of
(\ref{gamRt}) into (\ref{gammazeta}) leads to a divergent integrals. However
in this case a renormalization is not necessary, the integral in the r.h.s. of
(\ref{gamRt}) is expressed in terms of the error function
\[
Erf(z) = \frac{2}{\sqrt{\pi}} \int_{0}^{z} e^{-\tau^{2}}d\tau.
\]
namely, after some change of variables
\begin{equation}
\label{gamkt}\gamma_{k}(t) = Erf(b\sqrt{t}) = \frac{2b\sqrt{t}}{\sqrt{\pi}%
}\int_{0}^{1}\exp[-b^{2}t\tau^{2}]d\tau.
\end{equation}
The same result gives \cite{BE}. The Mellin transform of this representation
gives the following expression of zeta function of the operator L.
\begin{equation}
\label{zetaL}\zeta_{L}^{R}(s) = \frac{2b^{-2s}}{\sqrt{\pi}}\frac
{\Gamma(s+1/2)}{\Gamma(s)}\int_{0}^{1}\tau^{-2s-1}d\tau= -\frac{b^{-2s}}%
{\sqrt{\pi}}\frac{\Gamma(s+1/2)}{\Gamma(s+1)}.
\end{equation}
The integral in (\ref{zetaL}) converged only at $Re s <0$ but gives the
analytical continuation for all $s \in C$ excluding the poles of $\Gamma(s+1)$.

\begin{center}
\begin{figure}
  \includegraphics[width=4in]{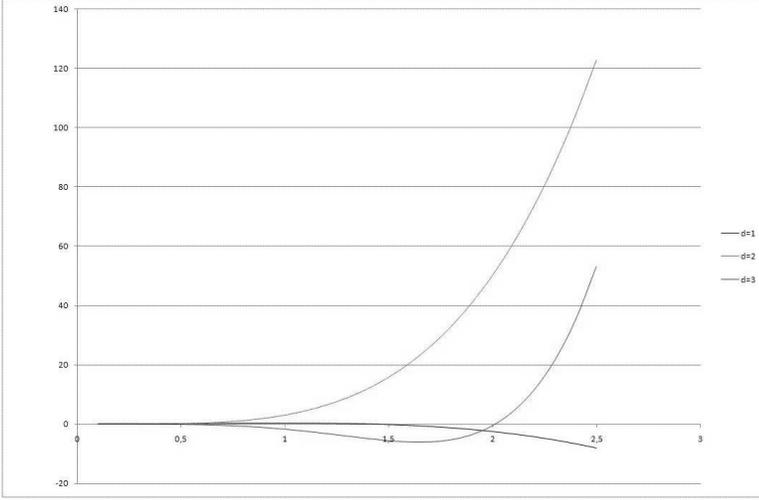}\\
      \caption{d=1,2,3 dependence of the derivative  $\frac{d\left(  \zeta_{L}^{D}(s)\right)  }{ds}$  to mass on the parameter m}\label{z}
\end{figure}
\end{center}

Substituting the result (\ref{zetaL}) into one arising from (\ref{gg}) with
account of (\ref{nu}) yields
\begin{equation}
\zeta_{L}^{D}(s)=-4(4\pi)^{\frac{d}{2}}m^{d-1-2s}\frac{\Gamma(s+1-d/2)}%
{(2s+1-d)\Gamma(s)}.\label{zetaLf}%
\end{equation}
The condition $b=m$ is taken into account. Differentiation of
(\ref{zetaLf}) by $s$ at the point $s=0$ gives the desired
correction
\begin{equation}
\begin{array}{c}
  -\frac{1}{2}\frac{d\left(  \zeta_{L}^{D}(s)\right)  }{ds}= \\
  -2(4\pi)^{\frac{d}{2}}%
m^{d-2s-1}\frac{\Gamma\left(  s-\frac{1}{2}d+1\right)
}{\Gamma\left( s\right)  \left(  2s-d+1\right)  ^{2}}\left(
\left(  d-2s-1\right)  \left(
\operatorname{Psi}\left(  -\frac{1}{2}d+s+1\right)  -2\ln m-\operatorname{Psi}%
\left(  s\right)  \right)  \ +2\right)  \label{final}
\end{array}
\end{equation}
next we plot the dependence of the correction $\frac{d\left( \zeta_{L}%
^{D}(0)\right) }{ds}$ on m (Fig \ref{z}).%
 We would remind about the choice of the constant  g=2 in the last
sections.

\section{Conclusion}

 The integrals in the elliptic  case are evaluated numerically by means of
rapidly converging series for theta-functions.

 It is known \cite{LeZa0} that it
is possible to form a periodic solution of a soliton model by a
shift operation in the complex plane of the soliton (kink)
parameter.

  In \cite{HKSF} the authors study the
diffusion of kinks. The Ginzburg-Landau (GL) model is very popular
in different aspects of solid state physics, e.g. for magnetics
\cite{Wint}. Some recent papers open new field for applications
\cite{B,MB,Paw}.

The investigation of dislocations dynamics by means of FK model
gives a direct possibility to check quantum soliton effects
 separating kink and elliptic solitons  contribution via the energy dependence on parameters \cite{HKSF}. It is
 interesting to incorporate our results in a real crystal
 thermodynamics   via statistical physics approach
 \cite{S} or a direct echo response evaluation  with correspondent measurements \cite{KTKY}.
In a review \cite{S} a quantization contribution is already
discussed. There also direct simulations  of kink-antikink pairs
\cite{WMor}.
\subsubsection{Acknowledgment} I  thank G.
Kwiatkowski for discussions and assistance in plotting of the SG
kinks mass dependence on $m$.

\end{document}